\documentclass[pre,preprint,nofootinbib,floatfix]{revtex4} 
\usepackage[dvips]{graphicx}
\usepackage{epsfig}
\usepackage{amssymb,amsfonts,amsmath}
\usepackage{bm}
\usepackage{subfigure}

\def\Itens{\mbox{\boldmath$I$\unboldmath}}

\def\qbold{\mbox{\boldmath $q$\unboldmath}}
\def\Dbold{\mbox{\boldmath $D$\unboldmath}}

\def\rbold{\mbox{\boldmath $r$\unboldmath}}

\def\ubold{\mbox{\boldmath $u$\unboldmath}}

\def\3dots{\:\raisebox{-0.5ex}{$\stackrel{\textstyle.}{:}$}\:}
\def\beq{\begin{equation}}
\def\eeq{\end{equation}}
\def\bea{\begin{eqnarray}}
\def\eea{\end{eqnarray}}

\begin{document}

\title{Redundancy and cooperativity in the mechanics of compositely crosslinked filamentous networks}

\author{Moumita Das} \affiliation{Department of Physics and Astronomy, Vrije Universiteit, Amsterdam, The Netherlands} 
\author{D. A. Quint} \affiliation{Department of Physics, Syracuse University, New York, United States}
\author{J. M. Schwarz} \affiliation{Department of Physics, Syracuse University, New York, United States}
\footnote{All authors contributed equally to this work.}

\begin{abstract}

The actin cytoskeleton in living cells has many types of crosslinkers. The mechanical interplay between these different crosslinker types is an open issue in cytoskeletal mechanics.
We develop a framework to study the  cooperativity and redundancy in the mechanics of filamentous networks with two types of crosslinkers: crosslinkers that allow free rotations of filaments and crosslinkers that do not. The framework consists of numerical simulations and an effective medium theory on a percolating triangular lattice. We find that the introduction of angle-constraining crosslinkers significantly lowers the filament concentrations required for these networks to attain mechanical integrity. This cooperative effect also enhances the stiffness of the network and suppresses non-affine deformations at a fixed filament concentration.  We further find that  semiflexible networks with only freely-rotating crosslinks are mechanically very similar to  compositely crosslinked flexible networks with both networks exhibiting the same scaling behavior.  We show that the network mechanics can either be redundant or cooperative depending on the relative energy scale of filament bending to the energy stored in the angle-constraining crosslinkers, and the relative concentration of crosslinkers. Our results may have implications for understanding the role of multiple crosslinkers even in a system without bundle formation or other structural motifs.  

\end{abstract}
\maketitle
\section{Introduction}
The mechanical response of most cells arises from the mechanics of its cytoskeleton, a polymeric scaffold that spans the interior of these cells, and its interaction with the extra-cellular environment.  The cytoskeleton is made up of complex assemblies of protein filaments crosslinked and bundled together by a variety of accessory proteins. For example, there are approximately 23 distinct classes of accessory proteins such as fascin, $\alpha$-actinin, and filamin A~\cite{kreis} that crosslink
filamentous-actin (F-actin),  a major component of the cytoskeleton
that is resposible for the mechanical integrity and motility of
cells. Given the multitude of crosslinkers, several natural questions arise: Are the different types of crosslinkers redundant, or do they each serve specific functions? Do they act independently or cooperatively? What are the consequences of their mechanics for the mechanical integrity and response of the cell?  
A mutation study of {\it dictyostelium discoideum} cells lacking a particular actin crosslinking can still grow, locomote, and develop, though with some defects, thereby suggesting at least partial redundancy in the crosslinker's mechanical function~\cite{rivero} .
On the other hand, two types of crosslinkers working cooperatively may produce enhanced mechanical response.  
This cooperativity has been demonstrated in stress fibers crosslinked with the actin binding proteins (ABP) $\alpha$-actinin and fascin, where stress fibers containing both $\alpha$-actinin and fascin were more mechanically stable than stress fibers containing only $\alpha$ -actinin or fascin~\cite{tseng}. In addition, it has been found that two different crosslinkers are required for actin bundle formation {\it in vivo}~\cite{tilney}.  It could also be the case that different crosslinkers work independently of one another such that the dominant crosslinker dictates the mechanical response of the network~\cite{schmoller}. Given these various possibilities, how the cell uses different crosslinking proteins to optimize for certain mechanical characteristics is an important open issue in cytoskeletal mechanics.
 
Here, we address this redundancy versus cooperativity issue by studying a model network of  semiflexible filaments crosslinked with two types of crosslinkers. We first study the mechanical properties of the model network with one type of crosslinker and then add the second type of crosslinker and look for mechanical similarities and differences with the original model network. In addition, we also address the redundancy versus cooperativity issue of two types of crosslinkers for networks made of flexible filaments. 

As for the two types of crosslinkers, we consider crosslinkers that allow the crossing filaments to rotate freely (freely-rotating crosslinks) and crosslinkers that constrain the angle between two filaments.  The ABP $\alpha$-actinin is a candidate for the former type of crosslinking mechanics: optical trapping studies demonstrate that two filaments bound by $\alpha$-actinin can rotate easily~\cite{courson}. As an example of the latter, we consider filamin A (FLNa), which binds two actin filaments at a reasonably regular angle of ninety degrees, suggesting that FLNa constrains the angular degrees of freedom between two filaments~\cite{nakamura}.  Here, we do not take into account the possible unfolding of FLNa since the energy to unfold filamin A is large~\cite{nakamura,didonna,gardel.filamin}, nor do we take into account the kinetics of FLNa since we seek to understand fully the mechanics in the static regime first.  There exist other possible examples of angle-constraining crosslinkers such as Arp2/3 that serves a dual role as an F-actin nucleator and a crosslinker~\cite{blanchoin}.  While its role as a nucleator has been emphasized in lamellipodia formation~\cite{pollard,svitkina}, its role constraining the angle between the mother and daughter filaments is presumably also important for lamellipodia mechanics.  Better understanding of the mechanical role of Arp2/3 in lamellipodia may also help to distinguish between the dendritic nucleation model for lamellipodia formation and a new model where Arp2/3 only nucleates new filaments but does not produce branches~\cite{urban}.  
 
In studying the mechanical properties of compositely crosslinked filamentous networks, we focus on the onset of mechanical rigidity as the filament concentration is increased above some critical threshold. This onset is otherwise known as rigidity percolation~\cite{degennes,feng.sen,feng,thorpe,thorpe.cai,sahimi,latva-kokko}.  Above this critical threshold, both experiments and theoretical studies of F-actin networks have observed distinct mechanical regimes. For dense, stiff networks the mechanical response is uniform or affine and the strain energy is stored predominantly in filament stretching modes. While for sparse, floppy networks one finds a non-affine response dominated by filament bending where the observed mechanical response of the network is inhomogeneous and highly sensitive to the lengthscale being probed~\cite{head,heussinger,wilhelm,gardel,das}. It has been recently reported that there exists a {\em bend-stretch} coupled regime for intermediate crosslinking densities and filament stiffnesses~\cite{broedersz}. 

While considerable progress has been made in understanding the mechanics of cytoskeletal networks that are crosslinked by one type of crosslinkers, compositely crosslinked networks are only beginning to be explored experimentally~\cite{schmoller,esue} as are composite filament networks with one type of crosslinker theoretically~\cite{huisman,das2}.

Here we investigate the mechanics of such networks as a function of the concentration and elasticity of the crosslinkers and the filaments.

\section{Model and Methods}
\begin{figure}
\begin{center}
\includegraphics[width=.4\textwidth]{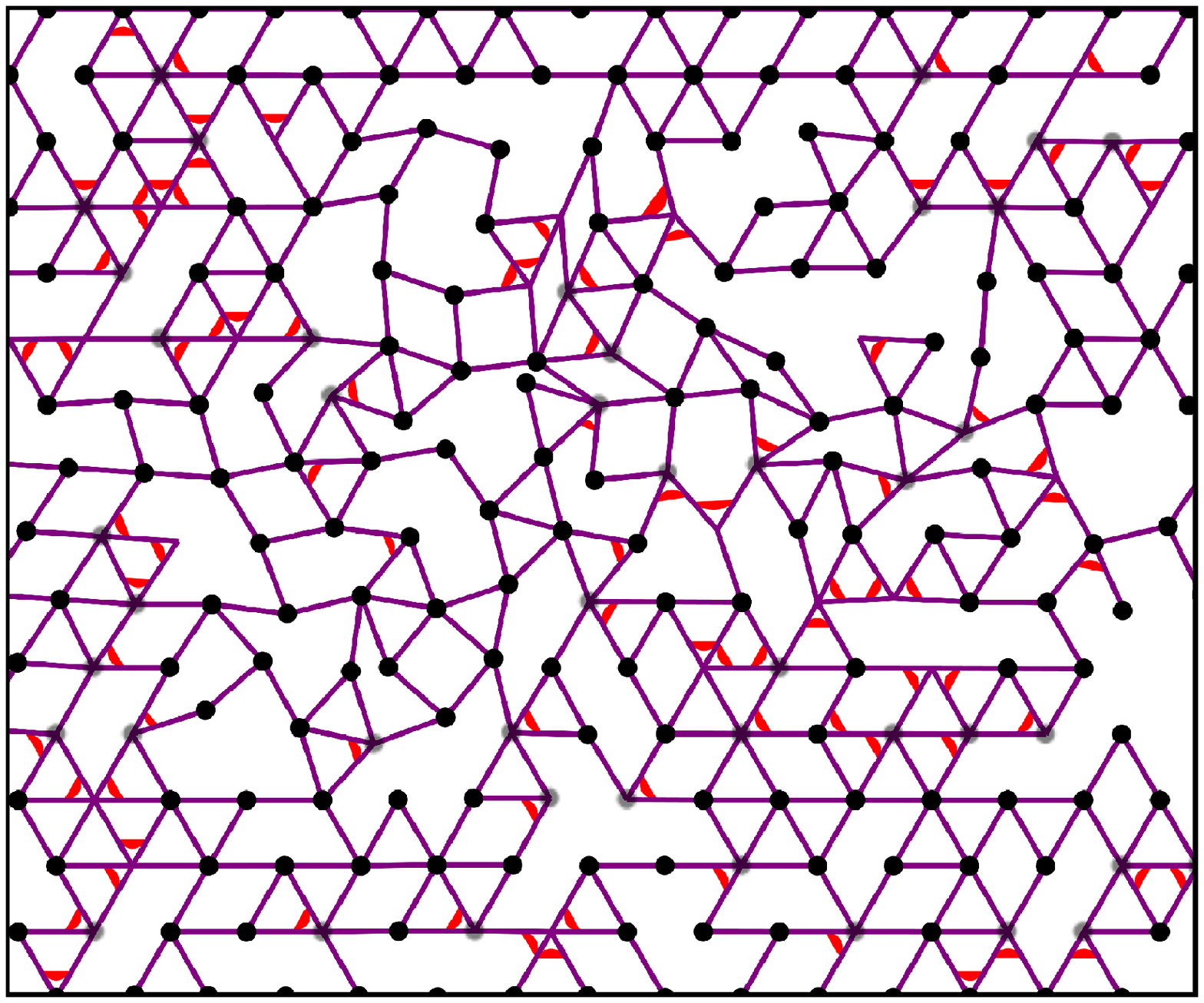}
\caption{Deformed configuration a compositely crosslinked semiflexible network with 2.7 percent strain, with bond occupation probability $p=0.64$, and angle-constraining crosslinker occupation probability $p_{nc}=0.15$ The purple lines denote semiflexible filaments, the red arcs denote angle-constraining crosslinks, the black circles represent nodes where all crossing filaments are free to rotate, while the grey circles denote nodes where some of the crossing filaments are free to rotate. The filament bending stiffness relative to stretching stiffness $\kappa/\alpha=10^{-6}$ and the stiffness of angular crosslinks relative to stretching stiffness $\kappa_{nc}/\alpha=10^{-6}$.}\label{Fig0}
\end{center}
\end{figure}

We arrange infinitely long filaments in the plane of a two-dimensional triangular lattice.  The filaments are given an extensional spring constant $\alpha$, and a filament bending modulus $\kappa$. We introduce finite filament length $L$ into the system by cutting bonds with probability $1-p$, where $0 < p < 1 $, with no spatial correlations between these cutting points. The cutting generates a disordered network with a broad distribution of filament lengths.  When two filaments intersect, there exists a freely-rotating crosslink preventing the two filaments from sliding with respect to one another. Next, we introduce angular springs with strength $\kappa_{nc}$ between filaments crossing at $60^\circ$ angles with a probability $p_{nc}$, where $nc$ denotes non-collinear.  These angular springs model the second type of crosslinker. See Fig.\ref{Fig0} for a schematic. 

We study the mechanical response of this disordered network under an externally applied strain in the linear response regime. 
For simplicity we set the rest length of the bonds to unity. Let  $\rbold_{ij}$ be the unit vector along bonds and $\ubold_{ij}=\ubold_{i}-\ubold_{j}$  the strain on the bond $ij$. 
For small deformation $u$,  the deformation energy is 

\bea \label{energies}
E \!\! &=& \!\!\!  \frac{\alpha}{2} \sum_{\langle ij \rangle}\! p_{ij}  \left (\ubold_{ij} . \rbold_{ij} \right)^2
+ \frac{\kappa}{2}\!\!\! \sum_{\langle \widehat{ijk} =\pi \rangle}\!\!\!\!\! p_{ij} p_{jk}  (\left( \ubold_{ji} + \ubold_{jk}) \times \rbold_{ji}  \right )^2 \nonumber \\
\!\!&+& \!\!\!   \frac{\kappa_{nc}}{2}\!\!\! \sum_{\langle \widehat{ijk} =\pi/3 \rangle}\!\!\!\!\!\!  p_{ij} p_{jk} \; p_{nc} \; {\Delta\theta_{ijk}}^2 
\eea
where $p_{ij}$ is the probability that a bond is occupied,  $\sum_{\langle ij \rangle}$ represents sum over all bonds and  $\sum_{\langle ijk \rangle}$ represents sum over pairs of  bonds sharing a node. The first term in the deformation energy corresponds to the cost of extension or compression of the bonds, the second term to the penalty for the bending of filament segments made of pairs of adjacent collinear bonds, and the last term to the energy cost of change in the angles between crossing filaments that meet at $60^\circ$ angle. Furthermore, for small deformations $\Delta\theta_{ijk} =  (\ubold_{ji} \times \rbold_{ji}  -   \ubold_{jk} \times \rbold_{jk}).(\rbold_{ji} \times \rbold_{jk}) =
 -\frac{(\ubold_{ji} . \rbold_{ji}  +  \ubold_{jk} . \rbold_{jk})}{2} + u_{ik}.\rbold_{ik} $. It is straightforward to see that the angular spring $\widehat{ijk}$ between $ij$ and $jk$ will contribute to an effective spring in parallel with $ik$, giving rise to an enhanced effective spring constant $\mu=\alpha + \frac{3}{2} \kappa_{nc}$ .

\subsection{Effective medium theory}
 
We study the effective medium mechanical response for such disordered networks following  the mean field theory developed in \cite{feng,thorpe} for central force networks and \cite{das} for filament bending networks. The aim of the theory is to construct an effective medium, or ordered network, that has the same mechanical response to a given deformation field as the depleted network under consideration. The effective elastic constants are determined by requiring that strain fluctuations produced in the original, ordered network by randomly cutting filaments  and removing angular springs vanish when averaged over the entire network. 

Let us consider an ordered network with each bond having a spring constant $\mu_m$,  a filament bending constant for adjacent collinear bond pairs $\kappa_m$, and an angular bending constant $\kappa_{nc,m}$ between bonds making $60^\circ$ angles. Under small applied strain, the filament stretching and filament bending modes are orthogonal, with stretching forces contributing only to deformations along filaments ($\ubold_{\parallel}$) and bending forces contributing only to deformations perpendicular to filaments ($\ubold_{\perp}$), and hence we can treat them separately. The angular forces due to the angular (non-collinear) springs, when present,  contribute to stretching of filaments as discussed earlier, where we only consider three body interactions. For these springs to contribute to bending one needs to consider four-body interactions which is outside the scope of this paper and will be addressed in future work. 

We start with the deformed network and replace a pair of adjacent collinear bonds with bending rigidity $\kappa_m$ by one with a rigidity $\kappa$, and a bond spring with extensional elastic constant $\mu_m$ by  a spring with an elastic constant $\mu$ and the facing $60^\circ$ angular spring by $\kappa_{nc}$.  This will lead to additional deformation of the above filament segments and the angle which we calculate as follows.
The virtual force that needs to be applied to restore the nodes to their original positions before the replacement of the bonds will have 
a stretching, a bending and an angular contribution: $F_s$, $F_b$, and $F_{\theta}$.  The virtual stretching force is given by 
$F_s = (\mu_m - \alpha - 3 \kappa_{nc}/2) {u}_{\parallel,m}$, the virtual filament bending force is  $F_b=  (\kappa_m -\kappa) {u}_{\perp,m}$, while the virtual force to restore the angle is  
$F_{\theta}=(\kappa_{nc,m} -\kappa_{nc}) {\theta_m}$, where ${u}_{\parallel,m}$, ${u}_{\perp,m}$ and ${\theta_m}$ are the corresponding deformations in the ordered network under the applied deformation field.  By the superposition principle, the strain fluctuations introduced by replacing the above bending hinges and bonds
 in the strained network are the same as the extra deformations that result when we apply the above virtual forces 
on respective hinges and segments in the unstrained network. The components of this ``fluctuation'' 
are, therefore, given by:

 \bea
d\ell_{\parallel} &=& \frac{F_s}{\mu_m/a^* - \mu_m + \alpha +  (3/2) \kappa_{nc}}  \nonumber \\
d\ell_{\perp} &=& \frac{F_b}{\kappa_m/b^*-\kappa_m + \kappa}   \nonumber \\
d{\theta}&=& \frac{F_{\theta}}{\kappa_{nc,m}/c^*- \kappa_{nc,m} + \kappa_{nc}} 
\eea

The effective medium spring and bending constants, $\mu_m$, $\kappa_m$ and $\kappa_{nc,m}$, respectively, can be calculated by demanding that
the disordered-averaged deformations $\langle {d\ell}_{\parallel} \rangle$, $\langle {d\ell}_{\parallel} \rangle$,  and $\langle d{\theta}  \rangle$  vanish, i.e. $ 
\left \langle \frac{\mu_m -\alpha  - 3 \kappa_{nc}/2}{\mu_m/a^*  -\mu_m + \alpha + 3 \kappa_{nc}/2} \right \rangle = 0$, 
$\left \langle \frac{\kappa_m -\kappa}{\kappa_m/b^*  -\kappa_m + \kappa}  \right \rangle = 0$, and 
$\left \langle \frac{ \kappa_{nc,m} - \kappa_{nc}}{\kappa_{nc,m}/c^*  - \kappa_{nc,m} + \kappa_{nc}} \right  \rangle =0$.
To perform the disorder averaging, since the stretching of filaments is defined in terms of spring elasticity of single bonds $\alpha$, the disorder in filament stretching is given by $P(\alpha')= p \delta(\alpha'-\alpha) + (1-p) \delta (\alpha')$. Filament bending, however, is defined on pairs of adjacent collinear bonds with the normalized probability distribution $P(\kappa') = p^2 \delta(\kappa'-\kappa) + (1-p^2) \delta(\kappa')$. Similarly,  for the angular springs, the normalized probability distribution is given by $P(\kappa'_{nc})=p_{nc} p^2  \delta(\kappa'_{nc} -\kappa_{nc}) + (1- p_{nc} p^2) \delta(\kappa'_{nc}))$.  This disorder averaging gives the effective medium elastic constants as a function of $p$ and $p_{nc}$ as  
\begin{eqnarray} 
\label{Triangular-EMT-Rp}
& &p^3 p_{arp} \left(\frac{\mu_m -\alpha  - 3 \kappa_{arp}/2}{\mu_m/a^*  -\mu_m + \alpha + 3 \kappa_{arp}/2}\right) 
+ (1-p) p^2 p_{arp} \left( \frac{\mu_m  - 3 \kappa_{arp}/2}{\mu_m/a^*  -\mu_m + 3 \kappa_{arp}/2} \right)\nonumber \\
& &\nonumber\\
&+& p (1-p^2 p_{arp}) \left( \frac{\mu_m  - \alpha}{\mu_m/a^*  -\mu_m + \alpha} \right)
+ (1-p)(1-p^2 p_{arp}) \left( \frac{\mu_m }{\mu_m/a^*  -\mu_m}\right) =0 \nonumber \\  
& &\nonumber\\
& & \frac{\kappa_m}{\kappa} = \frac{p^2-b^{*}}{1-b^{*}}, \hspace{5pt} \text{and}\hspace{5pt} 
 \frac{\kappa_{m,arp}}{\kappa_{arp}} = \frac{p_{arp} \;\; p^2-c^{*}}{1-c^{*}}.
\end{eqnarray}

The constants $a^{*}$, $b^{*}$ and $c^{*}$ for the network contribution to the effective spring constant $\mu_m/a^{*}$ of bonds, to the filament bending rigidity $\kappa_m/b^{*}$, and the bending rigidity  $\kappa_{nc}/c^{*}$ of angular springs making $60^\circ$ angles respectively, are  given by $a^{*}, b^{*}, c^{*} = \frac{2}{Nz} \sum_{q} Tr  \left[ \Dbold_{s,b,nc} (q) \Dbold^{-1} (q) \right ]$. The sum is over the first Brillouin zone and $z$ is the coordination number. The stretching, filament bending and non-collinear bending contributions, $\Dbold_{s,b,nc}(q)$ respectively,  to the full dynamical matrix $\Dbold(q)= \Dbold_s (q) + \Dbold_b(q) + \Dbold_{nc}(q)  $, are given by:

 \begin{eqnarray}
\label{Dmatrix-nc}
 \Dbold_s (q) &=& \mu_m \sum_{\langle ij \rangle}
\left[ 1 - e^{-i \qbold.\rbold_{ij}} \right]  \rbold_{ij}  \rbold_{ij}  \nonumber  \\
{\Dbold_b}(q) &=& \kappa_m  \sum_{\langle ij \rangle}  
 \left[ 4(1 - \cos(\qbold.\rbold_{ij}))  \right.  \nonumber  \\  && 
  \left. - ( 1 - \cos(2 \qbold.\rbold_{ij})) \right]   \left(\Itens-\rbold_{ij}  \rbold_{ij} \right) \nonumber \\
\Dbold_{nc} (q) &=& \frac{3}{2} \kappa_{nc,m} \sum \left[ 2 (1 - \cos(\qbold.\rbold_{ij})) + 2 (1 -
\cos(\qbold.\rbold_{ik})) \right.  \nonumber \\  &&
\left. - 2 (1 - \cos(\qbold.\rbold_{jk}))    \right] \rbold_{ij}  \rbold_{ik}\label{Dnc}
\end{eqnarray}
with $\Itens$ the unit tensor and the sums are over nearest neighbors ~\cite{feng}.
Note that for small $q$, $\Dbold_b \sim q^4$ and $\Dbold_s \sim q^2$
have the expected wavenumber dependencies for bending and stretching. 

By definition, $a^{*}+b^{*} +c^{*}=2d/z$, where $d=2$ is the dimensionality of the system.  
At the rigidity percolation threshold $p=p_{rp}$, $\mu_m$, $\kappa_m$ and $\kappa_{nc,m}$ vanish,
giving  $a^{*}=p+p^2 p_{nc} - p^3 p_{nc}$, $b^{*}=p^2$ and $c^{*}=p^2 p_{nc}$. 
For semiflexible filament networks with only freely-rotating crosslinks i.e. filament stretching and bending interactions only, 
the rigidity percolation threshold is given by  $p_{rp}=0.457$. For networks with angle-constraining crosslinks, at
$p_{nc}=1$, we obtain rigidity percolation thresholds $p_{rp}= 0.405$ for the case of flexible filament networks, and $p_{rp}=0.347$ for semiflexible filament networks. We also calculate how $p_{rp}$ changes on continuously increasing $p_{nc}$ from $0$ to $1$.

\subsection{Numerical Simulations}
Simulations were carried out on a triangular lattice with half periodic boundary conditions along the shear direction for the energetic terms whose small deformation limit is given in Eq. \eqref{energies}. Networks were constructed by adding bonds between lattice sites with probability $p$. Next, a shear deformation was applied to the two fixed boundaries of magnitude $\pm\gamma$. The lattice was then relaxed by minimizing its energy using the conjugate gradient method ~\cite{numrecipes} allowing the deformation to propagate into the bulk of the lattice. Once the minimized energetic state was found within the tolerance specified, in this case the square root of the machine precision $\sim 10^{-8}$, the shear modulus was then measured using the relation, $G=\frac{2E_{min}}{a_{cell}(\gamma L)^2}$, using small strains $<5\%$, with $L$ denoting the system length and $a_{cell}$ denoting the area of the unit cell for a triangular lattice which is equal to $3\sqrt{2}$ in our units.  System size $L=64$ was studied, unless otherwise specified, and sufficient averaging was performed.  

\section{Results}

\begin{figure}
\includegraphics[width=1.0\textwidth]{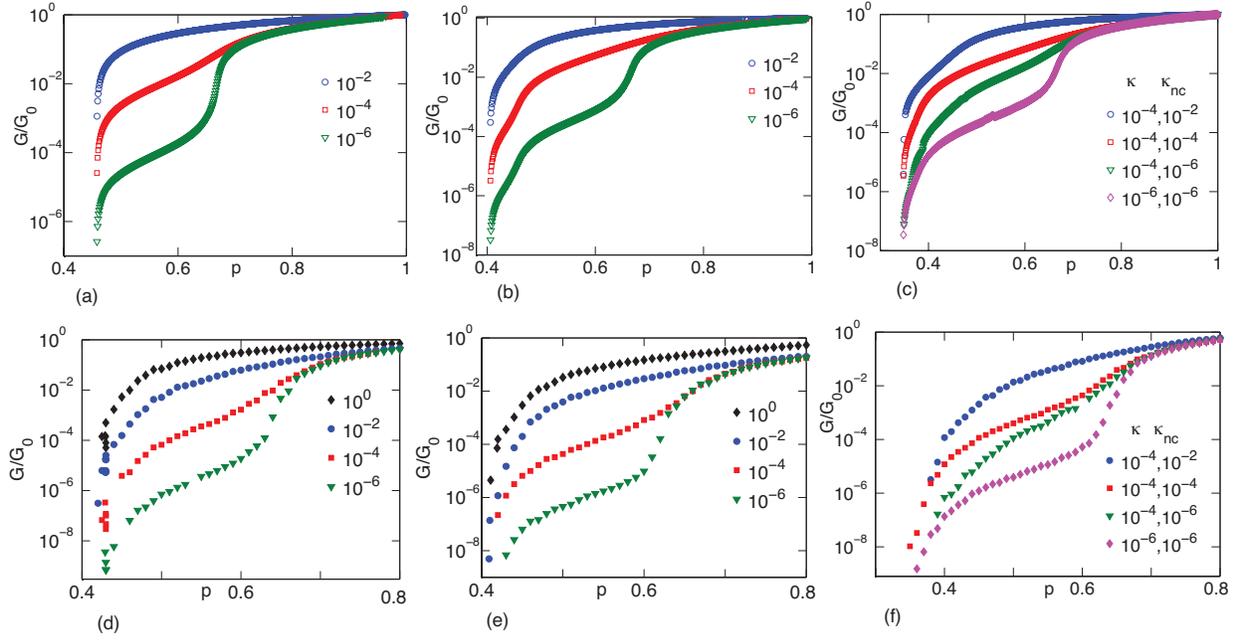}
\caption{The shear modulus as a function of $p$ for semiflexible networks with freely-rotating crosslinks ((a) and (d)), flexible networks with freely-rotating and angle-constraning crosslinks ((b) and (e)), and semiflexible networks with both crosslinkers ((c) and (f)).The top panels show results from the effective medium theory and bottom panels show results from the simulations.}\label{Fig1}
\end{figure}

\begin{figure}
\includegraphics[width=1.0\textwidth]{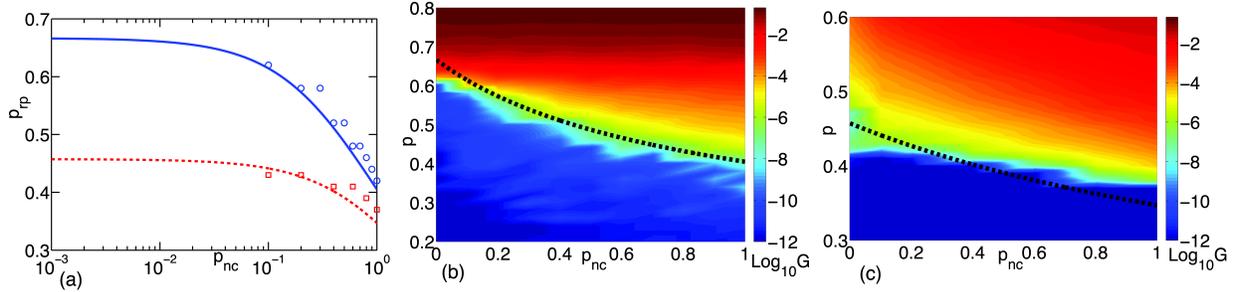}
\caption{The presence of angular constraints allows these networks to have a finite rigidity even for small concentration of filaments. Figure (a) shows how the rigidity percolation threshold can be continuously lowered by increasing the concentration of angular springs for flexible (solid, blue) and stiff (dashed, red) networks. The lines correspond to the effective medium theory and the symbols to the numerical simulation, where the system size $L=32$ for the semiflexible filaments and $L=64$ for the flexible filaments. Figures (b) and (c) show the shear modulus (in logarithmic scale described by the colorbar) as a function of $p$ and $p_{nc}$ for flexible networks (b) and semiflexible networks (c). The parameter values studied are (b) $\kappa_{nc}/\alpha=10^{-4}$ and (c) $\kappa/\alpha=10^{-4}$, $\kappa_{nc}/\alpha=10^{-2}$. The black dashed lines in (b) and (c) correspond to the effective medium theory prediction of the rigidity percolation threshold.
}\label{Fig2}
\end{figure}

{\bf Mechanical integrity as measured by the shear modulus:} On a triangular lattice, networks made solely of Hookean springs lose rigidity at a bond occupation probability around $p_{rp,I}=2/3$ ~\cite{maxwell,alexander,feng}. This result corresponds to the central force isostatic point at which the number of constraints is equal to the number of degrees of freedom on average. In contrast, networks made of semiflexible filaments become rigid at a smaller $p$ due to extra constraints placed on the system via filament bending. For semiflexible networks with freely-rotating crosslinks, our effective medium theory shows that the shear modulus, $G$, approaches zero at $p_{rp}=0.457$ as shown in Fig.\ref{Fig1} $(a)$. This result is in good agreement with our simulation results yielding $p_{rp}=0.442(6)$ and previous numerical results~\cite{broedersz}. See Fig.\ref{Fig1} $(d)$. A different formulation of the EMT yields $p_{rp}\approx 0.56$~\cite{broedersz}.  By introducing additional crosslinks that constrain angles between filaments at $60^\circ$, the rigidity percolation threshold is lowered.  Our EMT yields $p_{rp}=0.347$ and our simulations yield $p_{pr}=0.348(4)$ for $p_{nc}=1$ (Fig.\ref{Fig1} $(c)$ and $(f)$). The cooperative mechanical interplay between these crosslinks and their interaction with filaments allows the network to form a rigid stress-bearing structure at remarkably low crosslinking densities, almost immediately after it attains geometric percolation, $p_c=2\sin(\pi/18)$, which agrees with a calculation by Kantor and Webman~\cite{kantor}. For flexible filament networks, introducing angle-constraining crosslinkers also lowers the rigidity percolation threshold as compared to the isostatic point with the network attaining rigidity at $p_{rp}= 0.405$ for our EMT and $p_{rp}=0.408(4)$ in the simulations ((Fig.\ref{Fig1} $(b)$ and $(e)$). Incidentally, our result agrees very well with a previous simulation~\cite{hughes}. We also compute analytically and numerically how $p_{rp}$ changes with $p_{nc}$.  See Fig.\ref{Fig2}$(a)$, $(b)$ and $(c)$. Note that $p_{rp}$ is lowered continuously as the concentration of angle-constraining crosslinks is increased.

Just above the rigidity percolation threshold, for a semiflexible network with freely-rotating crosslinks, we find a bending-dominated regime for sparse networks with the shear modulus eventually crossing over to a stretch dominated affine regime at higher filament densities. The purely stretch dominated regime is represented by the macroscopic shear modulus $G$ staying almost constant with increasing $p$, while in the purely bend dominated regime the network is highly floppy and $G$ is a sensitive function of $p$, decreasing rapidly as $p$ is lowered. This behavior has been observed previously in ~\cite{head,heussinger,wilhelm,das,broedersz}. For $\kappa\ll\alpha$, both the effective medium theory and the simulations yield a bend-stretch coupled regime, which is characterized by an inflection in $G$ as a function of $p$ as observed most clearly for $\kappa=10^{-6}$ (with $\alpha=1$).  

We find a similar non-affine to affine crossover for the compositely crosslinked flexible filament networks and semflexible filament networks as $p$ is increased. For the flexible filament networks, however, the bend-stretch coupling regime occurs for $\kappa_{nc}\ll\alpha$, i.e. $\kappa_{nc}$ replaces $\kappa$. For semiflexible filament networks, as long as $\kappa_{nc}\lesssim \kappa<<\alpha$, the bend-stretch coupled regime is robust (for fixed $p_{nc}$). In contrast, for $\kappa<<\kappa_{nc}<<\alpha$, the angle-constraining crosslinker suppresses the bend-stretch coupled regime and enhances the shear modulus to that of an affinely deforming network (for fixed $p_{nc}$). The mechanics of the network has been altered with the introduction of the second type of crosslinker.

\begin{figure}
\includegraphics[width=.9\textwidth]{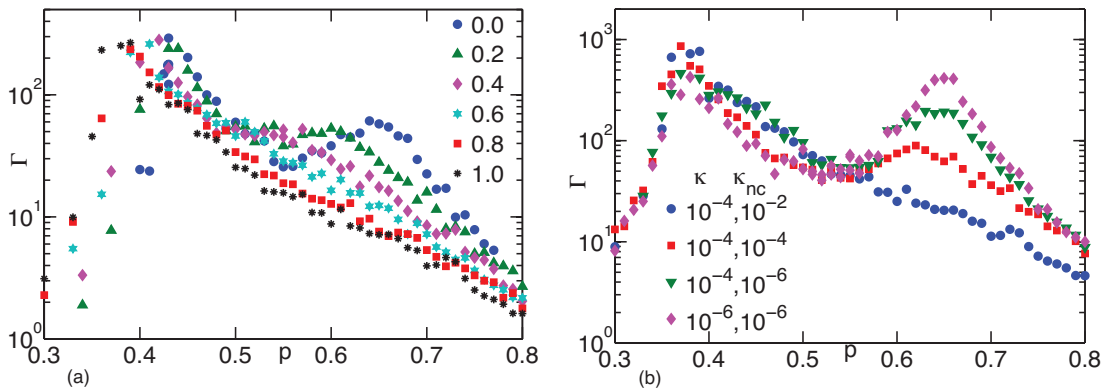}
\caption{The non-affinity parameter $\Gamma$ as a function of $p$ for semiflexible networks with both types of crosslinkers. In (a) we show the effect of changing the concentration $p_{nc}$ of the angle-constraining crosslinkers for $\kappa/\alpha=10^{-4}$, $\kappa_{nc}/\alpha=10^{-2}$ and $L=32$, while in (b) we show the effect of changing their stiffness $\kappa_{nc}$ for $L=64$.}\label{Fig3}
\end{figure}

{\bf Non-affinity parameter:} To further investigate how the interaction of the crosslinkers affects the affine and non-affine mechanical regimes, we numerically study a measure for the degree of non-affinity in the mechanical response, $\Gamma$, defined in Ref.\cite{broedersz} as: 
\beq
\Gamma=\frac{1}{L^2}{\gamma^2}\sum_{i}^{N}(\mathbf{u}_{i}-\mathbf{u}_{aff})^2.
\eeq\label{NA}
 The non-affinity parameter can be interpreted as a measure of the proximity to criticality, diverging at a critical point as we approach infinite system size. We find that $\Gamma$ develops a peak at the rigidity percolation threshold, which progressively moves to smaller values of $p$ as the concentration of angular crosslinkers $p_{nc}$ is increased (Fig.\ref{Fig3} $(a)$). A  second peak develops near the isostatic point for  $\kappa_{nc}\lesssim \kappa<<\alpha$ as seen in Fig.\ref{Fig3}  $(b)$.  As both the collinear and non-collinear bending stiffnesses tend to zero, the network mechanics approaches that of a central force network, and the second peak in $\Gamma$ at the isostatic point becomes increasingly more pronounced. 

On the other hand, this second peak can be suppressed by increasing $\kappa_{nc}/\kappa$ (Fig.\ref{Fig3} $(b)$), or by increasing the concentration $p_{nc}$ (Fig.\ref{Fig3} $(a)$) even for very small values of $\kappa/\alpha$. This further corroborates that adding angle-constraining crosslinkers to non-affine networks can suppress non-affine fluctuations, provided they energetically dominate over filament bending. The reason for this suppression can be understood by considering the effect of adding a constraint which prohibits the free rotation of crossing filaments. As the concentration of these non-collinear crosslinks $p_{nc}$ is increased (at fixed avg. filament length) microscopic deformations will become correlated. The lengthscale associated with this correlation will increase on increasing either $p$ or $p_{nc}$, and will eventually reach a lengthscale comparable to system size even at $p\sim p_{rp,I}$ at large enough concentration and/or stiffness of the angular springs.  As a result the mechanical response of the network will approach that of an affinely deforming network. Upon decreasing the value of $\kappa_{nc}/\alpha$ relative to $\kappa/\alpha$ we again recover the second peak because energetically the system can afford to bend collectively near the isostatic point.

\begin{figure}
\includegraphics[width=.9\textwidth]{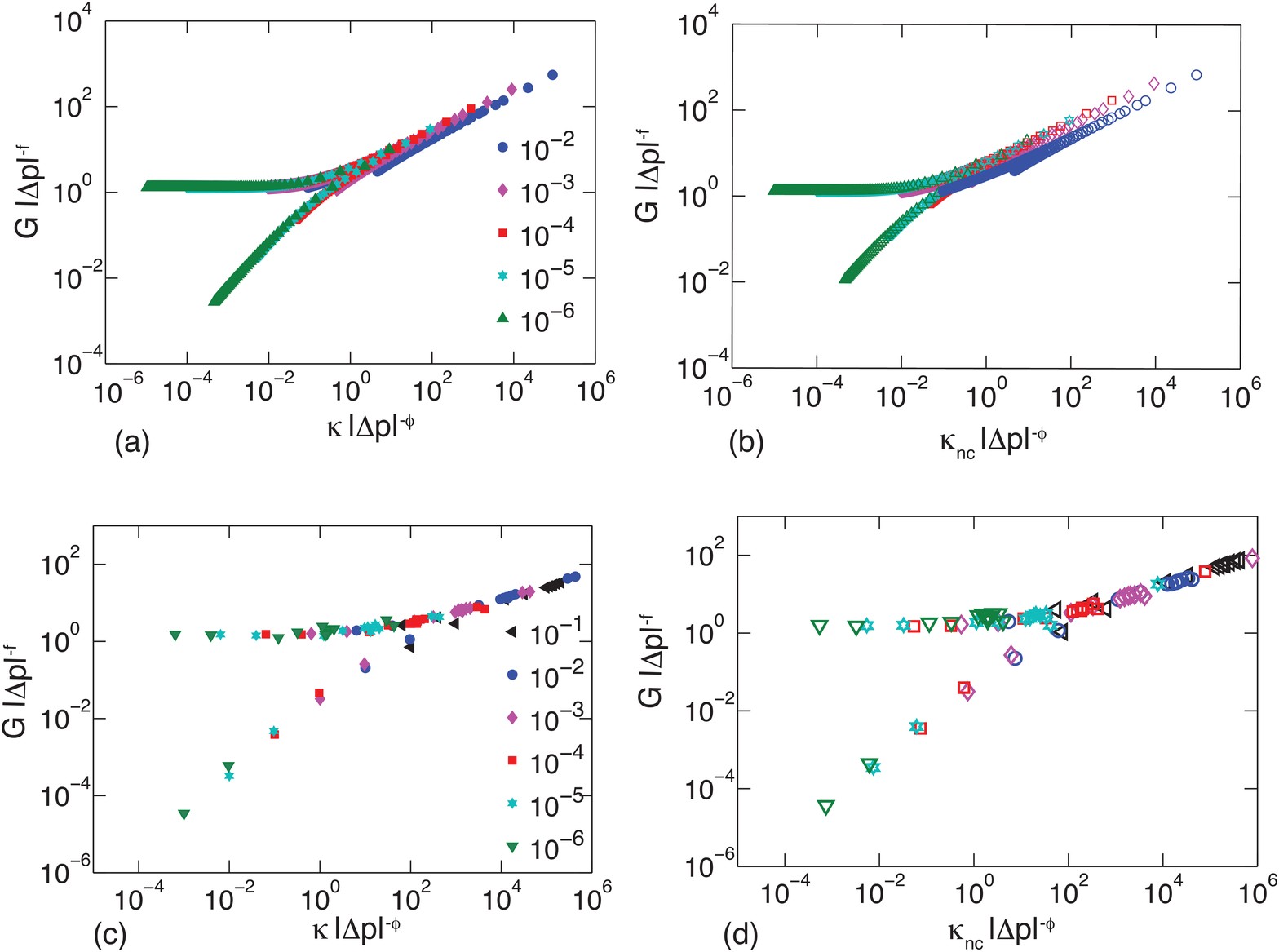}
\caption{Close to isostaticity, the shear modulus $G$ scales with $\Delta p= p-p_{rp,I}$  and $\kappa$ ($\kappa_{nc}$)  as  $G |\Delta p|^{-f} = \kappa |\Delta p |^{-\phi}$. The effective medium theory predicts mean field exponents $f=1$ and $\phi=2$ for both semiflexible networks with freely-rotating crosslinkers (a) and compositely crosslinked flexible networks (b), while simulations predict $f=1.1(1)$ and $\phi=2.9(1)$ for semiflexible networks with freely-rotating crosslinkers (c) and $f=1.1(1)$ and $\phi=2.8(1)$ for compositely crosslinked flexible networks (d).}\label{Fig4}
\end{figure}

{\bf Scaling near the isostatic point:} Finally, using scaling analysis we quantify the similarity in mechanics between freely-rotating crosslinked semiflexible networks and compositely crosslinked flexible networks. To do this, we examine the scaling of the shear modulus $G$ near the isostatic point with $\Delta p= p - p_{rp,I} \ll 1$. For $\kappa/\alpha \ll\Delta p$ (or $\kappa_{nc}/\alpha \ll \Delta p$), the shear modulus scales as $G=\alpha |\Delta p|^f \mathcal{G}_{\pm}(\frac{\kappa}{\alpha}|\Delta p|^{-\phi})$ (or $G=\alpha |\Delta p|^f \mathcal{G}_{\pm}(\frac{\kappa_{nc}}{\alpha}|\Delta p|^{-\phi})$)~\cite{broedersz,wyart}.  For both $(a)\,\kappa=0$,$\kappa_{nc} >0$ and $(b)\,\kappa > 0$, $\kappa_{nc} =0$, the EMT predicts $f=1$ and $\phi=2$ as shown in Fig.\ref{Fig4}(a) and (b), indicating that both types of networks demonstrate redundant, or generic, mechanics. To compare the EMT results with the simulations, we use the position in the second peak in $\Gamma$ to determine the central force percolation threshold, $p_{rp,I}$, and then vary $f$ and $\phi$ to obtain the best scaling collapse.  For case (a), $p_{rp,I}=0.666(3)$, $f=1.1(1)$ and $\phi=2.8(1)$.  For case (b), $p_{rp,I}=0.659(5)$, $f=1.1(1)$ and $\phi=2.9(1)$. Both sets of exponents are reasonably consistent with those found in Ref.~\cite{broedersz} for a semiflexible network with freely-rotating crosslinks only. Preliminary simulations for compositely crosslinked semiflexible networks indicate that the shear modulus scales as $G=\alpha |\Delta p|^f \mathcal{G}_{\pm}(\frac{\kappa}{\alpha}|\Delta p|^{-\phi},\frac{\kappa_{nc}}{\alpha}|\Delta p|^{-\gamma})$ also with a similar $f$ and a similar $\phi$ with $\phi=\gamma$ .

\section{Discussion}

In the limit of small strain, we conclude that the presence of multiple crosslinkers in living cells can be simultaneously cooperative and redundant in response to mechanical cues, with important implications for cell mechanics. Redundant functionality helps the cytokeleton be robust to a wide range of mechanical cues. On the other hand, different crosslinkers can also act cooperatively allowing the system to vary the critical filament concentration above which the cytoskeleton can transmit mechanical forces. 
This may enable the cytoskeleton to easily remodel in response to mechanical cues via the binding/unbinding of crosslinkers (tuning concentration) or their
folding/unfolding (tuning stiffness and type of crosslinker). Since the cytoskeleton consists of a finite amount of material, the ability to alter mechanics without
introducing major morphological changes or motifs may play important role in processes such as cell motility and shape change. 

{\bf Cooperativity:} In our study of two types of crosslinkers, crosslinkers that allow free rotations of filaments and crosslinkers that do not, we find two types of cooperative effects in the mechanics of such compositely crosslinked networks.  The first cooperative effect depends on the relative concentration of the two types of crosslinkers and second depends on the relative stiffness of the angle-constraining crosslinkers to the bending stiffness of the individual filaments. The first cooperative effect can be most strikingly observed beginning with an actin/$\alpha$-actinin network and increasing the concentration of FLNa, with $\alpha$-actinin representing the freely-rotating crosslinker~\cite{courson} and FLNa representing the angle-constraining crosslinker~\cite{nakamura}. By tuning the concentration of FLNa, the cell can modulate the minimum concentration of actin filaments necessary to attain mechanical rigidity, which 
can be essentially as low as the filament concentration required to form a geometrically percolating structure. This is in 
good agreement with the experimental observation that FLNa creates an F-actin network at filament concentrations lower than any other 
known crosslinker~\cite{nakamura}.  Increasing the FLNa concentration also suppresses the non-affine fluctuations near the rigidity percolation threshold by increasing the shear modulus of the network and giving rise to a more affine mechanical response while keeping the filament concentration fixed. 
Moreover, the cooperativity of $\alpha$-actinin and FLNa working to ehance the mechanical stiffness of actin networks has recently been observed in experiments~\cite{esue}. The addition of angle-constraining crosslinkers to flexible filament networks also decreases the concentration threshold required for mechanical rigidity, though the lower bound on the threshold is not as close as to geometric percolation as it is for semiflexible filaments. The lowering of the rigidity percolation threshold is independent of the energy scale of the crosslinker.  It depends purely on the number of degrees of freedom the crosslinker can freeze out between two filaments, i.e. the structure of the crosslinker. 

The second cooperative interplay between the two crosslinkers depends on the energy scale of the angle-constraining crosslinker to the filament bending energy. For $\kappa \ll \alpha$, the freely-rotating semiflexible filament system exhibits large non-affine fluctuations near the isostatic point.  Upon addition of the angle-constraining crosslinkers, for
$\kappa_{nc} \ge \kappa$, the non-affine fluctuations near this point become suppressed and the mechanics of the angle-constraining
crosslinker dominates the system. Once again, with a small change in concentration of the second crosslinker, the mechanical response of the network is changed dramatically.  

{\bf Redundancy:} We observe two redundant effects in these compositely crosslinked networks, the first of which depends on energy scales.  For $\kappa_{nc} \ll \kappa$ with $\kappa \ll \alpha$, the non-affine fluctuations near the isostatic point in the freely-rotating crosslinker semiflexible filament network remain large even with the addition of the angle-constraining crosslinker.  In other words, the angle-constraining crosslinkers are redundant near the isostatic point. Their purpose is to decrease the amount of material needed for mechanical rigidity as opposed to alter mechanical properties at higher filament concentrations.   

Redundancy is also evident in the mechanics of these networks sharing some important, generic properties.  All three networks studied here (free-rotating crosslinked semibflexible networks and compositely crosslinked semiflexible and flexible networks) have three distinct mechanical regimes: a regime dominated by the stretching elasticity of filaments, a regime dominated by the bending elasticity
of filaments and/or stiffness of angle-constraining crosslinkers, and an intermediate regime which depends on the interplay between these interactions.
The extent of these regimes can be controlled by tuning the relative strength of the above mechanical interactions. In particular, the ratio of bending rigidity to extensional modulus of an individual actin filament is $\sim 10^{-3}$ \cite{head}.  Since the bend-stretch coupled regime has not been observed in prior experiments on {\it in-vitro} actin networks crosslinked with FLNa only, we conjecture that the energy cost of deformation of angles between filaments crosslinked with FLNa is larger than the bending energy of filaments. The qualitative redundancy becomes quantitative, for example, near the isostatic point where  
we obtain the same scaling exponents for $G$ as a function of $p-p_{rp,I}$ and $\kappa$(or $\kappa_{nc}$) for the free-rotating crosslinked semiflexible network and the compositely crosslinked flexible network. Preliminary data suggests the same scaling extends to compositely crosslinked semiflexible networks. This result is an indication of the robustness of these networks and should not be considered
as a weakness. Whether or not this robustness extends to systems experiencing higher strains such that nonlinearities emerge is not yet known.

{\bf Lamellipodia mechanics:} The interplay between cooperative and redundant mechanical properties may be particularly important for the mechanics of branched F-actin networks in lamellipodia. Within lamellipodia, there exist some filament branches occuring at an angle of around $70^\circ$ with respect to the plus end of the mother filament (referred to as $Y-$ junctions). These branches are due to the ABP Arp2/3~\cite{blanchoin}.  During lamellipodia formation, these branches are presumed to be the dominant channel for filament nucleation. The mechanics of Arp2/3 can be modeled as an angular spring between the mother and daughter filament with an angular spring constant of approximately $10^{-19} J \,rad^{-2}$~\cite{blanchoin}.  In other words, Arp2/3 is an angle-constraining crosslinker for $Y-$junctions (as opposed to $X-$junctions), and thereby plays an important role in lamellipodia mechanics as demonstrated in this work.  The mechanical role of Arp2/3 in lamellipodia has not been investigated previously and may help to discriminate between the dendritic nucleation model~\cite{pollard,svitkina} and a new model~\cite{urban} by predicting the force transmitted in lamellipodia as a function of the Arp2/3 concentration. 

In addition to Arp2/3, FLNa localizes at $X-$junctions in the lamellipodia and is thought to stabilize the dendritic network \cite{revenu}.  Both angle-constraining crosslinkers lower the filament concentration threshold required for mechanical rigidity in the system. Depending on the energy scale of FLNa as compared to the energy scale of Arp2/3, addition of the FLNa may or may not modulate, for example, the bend-stretch coupling regime at intermediate filament concentrations. Again, at times mechanical redundancy is needed and at times not.  With three crosslinkers, the system can maximize the redundancy and the cooperativity. Of course, lamellipodia are dynamic in nature and are anisotropic since the Arp2/3 is activated from the leading edge of a cell. Both attributes will modulate the mechanical response. 

{\bf Outlook:} 
We have demonstrated both cooperativity and redundancy in the mechanics of compositely crosslinked filamentous networks. We have done so while maintaining the structure of an isotropic, unbundled filament network. Of course, crosslinkers can alter the morphology of the network via bundling, for example. In other words, different crosslinkers serve specific functions.  This specificity results in a change in microstructure. This will presumably affect the mechanics such that the cooperative and redundant interactions between multiple crosslinkers may differ from the above analysis. For example, the crosslinker that dominates in terms of creating the morphology will presumably dominate the mechanics. Schmoller and collaborators~\cite{schmoller} suggest that crosslinker with the higher concentration determines the structure and, therefore, the mechanics. Instead of redundancy or cooperativity, the specificity leads to the simple additivity of two types of crosslinkers in that different crosslinkers act independently of one another. In this study, however, we find both cooperativity and redundancy in the network mechanics even in the absence of such structural changes~\cite{wagner}, which, is arguably less intuitive and, therefore, more remarkable. Finally, while our focus here has been on the actin cytoskeleton as an example of a filamentous network, our results can be extended to collagen networks as well~\cite{stein}.

\begin{acknowledgments}
DAQ would like to thank Silke Henkes and Xavier Illa for useful discussions regarding lattice simulations. MD would like to thank Alex J. Levine, F. C. MacKintosh, C. Broedersz, T.C. Lubensky, C. Heussinger and A Zippelius for discussions on the mechanics of semiflexible networks. MD and JMS also acknowledge the hospitality of the Aspen Center for Physics where some of the early discussions took place. JMS is supported by NSF-DMR-0654373. MD is supported by a VENI fellowship from NWO, the Netherlands.
\end{acknowledgments}

\newpage

\end{document}